\documentclass[aps,prb,
twocolumn,superscriptaddress,amsmath,amssymb,verbatim]{revtex4-2}
\usepackage{amsfonts,stmaryrd,wasysym,graphicx,multirow,textcomp,subfigure,makecell}
\usepackage[font=small,labelfont=bf, justification=justified, format=plain]{caption} 

\usepackage{url}

\usepackage{dcolumn}
\usepackage{bm}
\usepackage[colorlinks=true,citecolor=blue,urlcolor=blue]{hyperref}
\hypersetup{colorlinks=true, urlcolor=blue, citecolor=cyan, pdfborder={0 0 0},}
\usepackage{xcolor}

\def\be{\begin{equation}}
        \def\ee{\end{equation}}
\def\bea{\begin{eqnarray}}
        \def\eea{\end{eqnarray}}

\renewcommand{\Re}{\mathrm{Re}\,}
\renewcommand{\Im}{\mathrm{Im}\,}

\newcommand{\p}{\partial}

\renewcommand{\Re}{\mathrm{Re}\,}
\renewcommand{\Im}{\mathrm{Im}\,}

\newcommand{\bs}{\boldsymbol}
\DeclareMathAlphabet{\bi}{OML}{cmm}{b}{it}

\def\ba{\begin{aligned}}
\def\ea{\end{aligned}}
\def\be{\begin{equation}}
\def\ee{\end{equation}}
\def\bearr{\begin{eqnarray}}
\def\eearr{\end{eqnarray}}
\def\rv{{\bf r}}
\def\kv{{\bf k}}
\def\dv{{\bf d}}

\begin{document}


\title{Non-linear spin current in the surface states of topological insulators}

\author{Srijan Chatterjee}
\thanks{Corresponding author: \href{mailto:srijanc24@iitk.ac.in}{srijanc24@iitk.ac.in}}
\affiliation{Department of Physics, Indian Institute of Technology Kanpur, Kanpur 208 016, India}

\author{Tarun Kanti Ghosh}
\affiliation{Department of Physics, Indian Institute of Technology Kanpur, Kanpur 208 016, India}

\date{\today}

\begin{abstract}
Spin Hall effect is one of the primary sources of pure spin current in spintronic devices, observed in spin-orbit coupled materials. 
In this work, we investigate linear and non-linear pure spin currents in the hexagonally warped surface states of topological insulators like $\rm Bi_2Te_3$, using the formalism of conserved spin transport.
The hexagonal warping effect enables the application of this formalism by providing a well-defined Berry curvature without breaking the time-reversal symmetry.
The absence of in-plane spin currents in the linear response motivates the extension of the formalism to the non-linear regime, by combining the perturbation of Bloch states with the Boltzmann transport equation. 
The second-order response exhibits both intrinsic and extrinsic in-plane spin currents induced by the spin-weighted Berry curvature, Berry connection polarizability (BCP) dipole and Berry curvature dipole (BCD) respectively.
The band anisotropy due to the $C_{3v}$ symmetry of the material ($\rm Bi_2Te_3$) propagates through the spin textures and the band geometric quantities, resulting in the selection of certain components of the spin conductivity tensors. 

\end{abstract}

\maketitle

\section{Introduction}
The transport and manipulation of the electronic spin is indispensable for building energy efficient spintronic devices\cite{Intro_1,spintronics_book,Fabian}. A pure spin current, primarily generated by the spin Hall effect (SHE) \cite{sinova2004universal,murakami2003dissipationless}, leads to the transport of spin without any Joule heating, in principle.  
Apart from its significance in spintronics, the physics of the spin Hall effect is also very rich\cite{SHE_Review, Gauge_physics_SHE}. It belongs to the class of anomalous Hall effect since it does not require an external magnetic field but more importantly, neither does it require the time-reversal symmetry($\mathcal{T}$) to break.  
Preserving the time-reversal symmetry is crucial to generate a dissipation-less pure spin current, by suppressing an accompanying charge current.

Spin Hall effect is generally observed in spin-orbit coupled semi-conductor devices\cite{SHE_Kato_exp}, the common examples being two-dimensional electron\cite{Schliemann-Loss-2DEG,sinova2004universal} and hole gases\cite{Schliemann-Loss-2DHG,Bernevig_holegas} with Rashba and Dresselhaus spin-orbit coupling\cite{Rashba-SOC,Dresselhaus-SOC}. 
In these systems, the commonly used conventional spin current\cite{sinova2004universal}, defined as the symmetrized product of the velocity and spin operators, is not conserved due to spin-orbit coupling. There is an additional spin torque term in the spin continuity equation which can be written as the divergence of spin torque dipole. This has been used to define a conserved spin current by Niu and others\cite{Niu-conserved-current,Niu-proper-spin-current} which is essential to explain spin accumulation at the sample boundaries due to a bulk spin current. 
Contrary to the conventional spin current operator that ascribes SHE to the spin Berry curvature\cite{spinQGT}, the conserved spin current reveals the Berry curvature as the source of SHE\cite{Niu-conserved-current,Culcer-proper_spin_current}. The role of Berry curvature in SHE has been  explored explicitly in Ref. \cite{Priyadarshini_kapri,Culcer-proper_spin_current} and it can be  extended to the non-linear regime which forms the central goal of this article.  

The Berry curvature, however, vanishes at each point in the momentum ($\kv$) space (except $\kv=0$) for most $\mathcal{T}$-symmetric two-dimensional two-level systems. Such systems are modelled by Hamiltonians of the form, $H = d_x(\kv)\sigma_x + d_y(\kv)\sigma_y$, where $\sigma_i$ are the Pauli matrices and $\dv(\kv)$ vanishes at $\kv =0$, resulting in a band-touching point where the Berry curvature shows a singularity\cite{Fujita_2010, Chang}.  
In order to have a finite and well-behaved Berry curvature at all $\kv$-points, one can introduce a mass term ($m\sigma_z$) and open a gap at $\kv=0$. Spin Hall effect due to Berry curvature in a gapped Rashba model has been studied in Ref.\cite{Priyadarshini_kapri}, where the spin Hall conductivity for the gap-less Rashba model\cite{sinova2004universal} has also been reproduced by taking the limit $m\rightarrow0$. However, this comes at the expense of breaking time-reversal symmetry which is not suitable for a pure spin current.

The effective two-band model for the surface states of 3D topological insulators ($\rm{Bi_2Te_3}, \rm{Bi_2Se_3}$, etc.)\cite{Zhang-Nature,model-Hamiltonian-SCZhang} provides a $\mathcal{T}$-preserved Hamiltonian that leads to a well-defined Berry curvature \cite{Awadesh, TanayNag}. This is due to a $k^3$-hexagonal warping term \cite{Fu-hexagonal-warping} that appears as a correction to the linear Dirac Hamiltonian in the low energy regime, near the band-touching point.  
It can be thought of as a counterpart of the anisotropic Dresselhaus spin-orbit interaction\cite{conserved_spin_current_holegas,Ojasvi}, with the important distinction that it enters the diagonal part of the Hamiltonian, leading to a non-vanishing Berry curvature and out-of-plane spin polarization.
The effect of hexagonal warping on the transport properties\cite{Effect_hex_wrap_transport,surface_charge_cond} and optical conductivity\cite{Carbotte} of the surface states have been studied extensively.
These 3D topological insulators are also expected to serve as efficient materials for spintronics\cite{spintronics_hex_warp}, showing large spin transfer torque\cite{large_STT_TI}, bulk and surface spin conductivity\cite{Bulk_surface_spin_cond,SHE_bulk_TI-Culcer} and spin galvanic effect\cite{spin-extraction-TI}.
Recently, the non-linear anomalous Hall conductivity\cite{Awadesh,TanayNag,nonlinear-Bi2Te3} in the hexagonally warped surface states of $\rm{Bi_2Te_3}$ has been explored theoretically, showing the third-order Hall current to be of leading order.
The suppression of the anomalous Hall current up to second-order\cite{TanayNag} opens an avenue to obtain pure spin current from these surface states.  

With this motivation, we investigate linear and second-order spin Hall effect in the surface states of 3D topological insulators. 
The presence of hexagonal warping makes it possible to apply the formalism of Berry curvature-induced conserved spin current directly, without adding an extra mass term.
Further, the article focusses on the orientation of spins relative to the direction of the spin current and reveals both in-plane and out-of-plane spin currents, including collinearly polarized spin current (CPSC)\cite{CPSC-wang}.

The rest of the article is organized as follows: section(\ref{sec:Formalism}) explains the formalism of non-linear spin current using first-order perturbation of eigenstates and the Boltzmann transport equation. Section(\ref{sec:surface-states}) presents the model Hamiltonian for the surface states and calculates the linear and second-order spin conductivity (SOSC), followed by a detailed analysis of the results. Further discussions are presented in section(\ref{sec:discussion}).

\section{Formalism}\label{sec:Formalism}
\subsection{Non-linear spin current}
The formalism for conserved spin transport has been developed in Ref. \cite{Niu-conserved-current, Culcer-proper_spin_current}, which ensures that circulating spin currents vanish at equilibrium \cite{Niu-proper-spin-current} and the Onsager's reciprocity relations are obeyed \cite{planar-SHE}.
This has already been applied to Dirac-Rashba systems \cite{Dirac-Rashba}, revealing linear intrinsic spin Hall current. 
Following this formalism, the $\kv$-resolved spin current due to an applied electric field ${\bf E}$, for the $n^{\rm th}$ band is given by
\begin{equation}\label{spin-current}
     j^l_{a,n} = e\Omega^{ab}_n S^l_n E_b,
\end{equation}
where $\Omega^{ab}_n$ and $S^l_n=\langle n\vert \hat S\vert n \rangle$ are the Berry curvature and the spin polarization in the $n^{\rm th}$ band, respectively. 

Considering the transport to be restricted on the $xy$-plane, the spin currents can be classified as `out-of-plane' $(l=z)$ and `in-plane' $(l=x,y)$. Reference \cite{Dirac-Rashba}, however, reports only an out-of-plane spin current and in-plane spin currents remain largely unexplored in 2D systems.  
In order to find these in-plane spin currents, we extend Eq. (\ref{spin-current}) to second-order in ${\bf E}$, by introducing perturbative corrections to the eigenstates ($\vert n\rangle$) and the equilibrium Fermi distribution $f_n(\kv)$.  
The first-order correction to the eigenstates due to an applied electric field is given by
\begin{equation}
    \vert n^{(1)}\rangle = \sum_{m\neq n} \frac{-e{\bf E}\cdot{\bf A}_{mn}(\kv)}{\varepsilon_{mn}}\vert m^{(0)}\rangle,
\end{equation}
where $\varepsilon_{mn}= \left(\varepsilon_m^{(0)}-\varepsilon_n^{(0)} \right)$ with $\varepsilon_m^{(0)}$ and $\vert m^{(0)}\rangle$ being the unperturbed energy eigenvalues and eigenstates, respectively. 
The inter-band Berry connection ${\bf A}_{mn}(\kv)$ arises from the off-diagonal components of the position operator that appears in the perturbation, $H'=-e{\bf E}\cdot\rv$.

Now in Eq. (\ref{spin-current}), the corrections can be introduced at two points: first, the correction to the eigenstates $\vert n\rangle$ in the expectation value of $\hat S^l$ and second, the correction to the Berry curvature $\Omega^{ab}_n$.  
Up to first-order in ${\bf E}$, the expectation value of $\hat S^l$ can be expanded as,
\begin{align}\label{modified-S}
   S'^l_n= \langle n' \vert\hat S^l \vert n'\rangle 
    \simeq S^l_n + 2\Re\{\langle n^{(1)}\vert \hat S^l \vert n^{(0)}\rangle\},
\end{align}
where $\vert n'\rangle = \vert n^{(0)}\rangle + \vert n^{(1)}\rangle$.
Further, the intra-band Berry connection evaluated using the first-order corrected eigenstates becomes \cite{kaplan}
\begin{equation}\label{bcp}
    A'^a_n = A^a_n - eG^{ab}_nE_b,
\end{equation}
where $G^{ab}_n$ is the Berry connection polarizability (BCP), given by
\begin{equation}
    G^{ab}_n = -2\Re\left\{ \sum_{m\neq n}\frac{ v^a_{nm} {v}^b_{mn}}{\varepsilon_{mn}^3}\right\}.
\end{equation}
Here, $ \hat v^a_{mn}$ are the matrix elements of the band velocity.
Using Eq. (\ref{bcp}), the first-order correction to the Berry curvature is found to be,
\begin{equation}\label{modified-BC}
    \Omega'^{ab}_n = \Omega^{ab}_n - e\partial_a G^{bc}_nE_c + e\partial_bG^{ac}_n E_c.
\end{equation}
Now, from Eq. (\ref{spin-current}) the corrected spin current for the $n^{\rm th}$ band is given by,
\begin{equation}
    j'^l_{a,n} = e\Omega'^{ab}_n S'^l_nE_b .
\end{equation}
Using equations (\ref{modified-S}) and (\ref{modified-BC}), the first-order spin current is given by
\begin{equation}\label{1st-order}
    j^{l,(1)}_{a,n} = e\Omega^{ab}_n S^l_n E_b,
\end{equation}
while the second-order spin current is obtained in the form,
\begin{equation}\label{2nd-order}
    j^{l,(2)}_{a,n} = \left( \Pi^{l,(\rm BCP)}_{abc,n} + \Pi^{l,(\rm BC)}_{abc,n} \right) E_bE_c,
\end{equation}
where the first term is a BCP dipole induced spin current, given by
\begin{equation}\label{BCP-term}
    \Pi^{l,(\rm BCP)}_{abc,n} = \frac{e^2}{4}\left[\partial_bG^{ac}_n + \partial_cG^{ab}_n - 2\partial_aG^{bc}_n\right]S^l_n, 
\end{equation}
and the second term is proportional to the Berry curvature,
\begin{equation}\label{BC-term}
    \Pi^{l,(\rm BC)}_{abc,n} = e^2\Omega^{ab}_n \: \Im \left\{ \sum_{m\neq n} \frac{S^l_{mn}v^c_{mn}}{\varepsilon^2_{mn}} \right\}.
\end{equation}
The total spin current is given by
\begin{equation}
    J^l_a = \sum_n\int_\kv [d{\bf k}] j'^l_{a,n} f_n(\kv),
\end{equation}
where $[d{\bf k}] = d^2k /(2\pi)^2$ and $f_n(\kv)$ is the Fermi-Dirac distribution for the $n^{\rm th}$ band, which is driven out of equilibrium by the applied electric field. Within the relaxation time approximation \cite{Ashcroft76},  $f_n(\kv) $ can be obtained by solving the Boltzmann transport equation
\begin{equation}
     \frac{e}{\hbar}{\bf E} \cdot {\bs \nabla}_{\bf k} f_n({\bf k}) = 
 - \frac{[f_n({\bf k}) - f_{n}^{(0)}({\bf k}) ]}{\tau},
\end{equation}
where $ \tau $ is the relaxation time of the charge carriers and $f_n^{(0)}({\bf k})= 1/(e^{[\epsilon_n({\bf k})-\mu]/(k_BT)} +1)$ is the equilibrium distribution in absence of the electric field.
The iterative solution for $f_n({\bf k}) $ can be written as
\begin{equation}
f_n({\bf k}) = \sum_{p=0,1,2,...}\left(\frac{e\tau}{\hbar}\right)^{p} 
\frac{\partial^{p}f_n^{(0)}}{\partial k^{p}_{i}}E^{p}_{i}.
\end{equation}
The total modified spin current is given by
\begin{equation}
     J^{l}_{a} = \sum_{n,p}\int_{\kv} [d \kv] j'^{l}_{i,n}\left(\frac{e\tau}{\hbar}\right)^{p}\frac{\partial^{p}f_n^{(0)}}{\partial k^{p}_{i}}E^{p}_{i}.
\end{equation}
Using equations (\ref{1st-order}) and (\ref{2nd-order}), the first and second-order contributions to the total spin current can be written as 
\begin{equation}
    J_a^l = \chi^l_{ab}E_b + \Gamma^l_{abc}E_bE_c,
\end{equation}
where the linear conductivity is given by
\begin{equation}\label{1st-order-cond}
    \chi^l_{ab} = \frac{e}{2}\sum_n\int_{\kv} [d\kv] \Omega^{ab}_n S^l_n f_n^{(0)}(\kv),
\end{equation}
and the SOSC is given by
\begin{align}\label{2nd-order-spin-cond}
    \Gamma^l_{abc} = &\frac{e^2\tau}{2\hbar}\sum_n \int_\kv [d\kv]\Omega^{ab}_n S^l_n\frac{\p f^{(0)}_n}{\p k_c} \notag\\
&+\sum_n\int_\kv [d\kv]\left( \Pi^{l,(\rm BCP)}_{abc,n} + \Pi^{l,(\rm BC)}_{abc,n} \right) f_n^{(0)}(\kv).
\end{align}

\subsection{Results for a generic two-level system}
The generic Hamiltonian for a two-level system is given by
\begin{equation}\label{2level-Ham}
     H(\kv) = \dv(\kv)\cdot{\bm\sigma},
\end{equation}
where the components of $\dv$ are functions of $\kv$ and the energy dispersion is given by $\varepsilon(\kv) = \pm\vert \dv(\kv)\vert=\pm d$.
The Berry curvature for two-level systems is given by\cite{review-articleQGT}
\begin{equation}\label{BC}
    \Omega^{ab}_\pm = \mp\frac{1}{2} \frac{\dv\cdot(\partial_a\dv\times\partial_b\dv)}{d^3}.
\end{equation}
The Berry connection polarizability (BCP) can be derived from the quantum metric as\cite{review-articleQGT}, $G^{ab}_n=-\frac{\p g^{ab}_n}{\p \varepsilon_n}$, $g^{ab}_n$ being the quantum metric. For a two-level system the BCP simply becomes 
\begin{equation}\label{BCP}
    G^{ab}_\pm= \pm\frac{\hbar^2}{4d}g^{ab}_\pm,
\end{equation}
where the quantum metric is given by\cite{review-articleQGT}
\begin{equation}
    g^{ab}_\pm = \frac{1}{4 d^2}\left[\p_a\dv\cdot\p_b\dv - \frac{1}{d^2}(\p_a\dv\cdot\dv)(\p_b\dv\cdot\dv) \right].
\end{equation}
Assuming that the spin operators are represented by Pauli matrices (${\hat{\bf S}}\propto{\hat{\bm\sigma}}$) and fixing the electric field along the $x$ direction, the BCP-dependent $\kv$-resolved spin conductivity from Eq. (\ref{BCP-term}) becomes
\begin{equation}\label{bcp-term-2-level}
    \Pi^{l,\rm (BCP)}_{y,\pm} = \frac{e^2}{2}\left[\p_xG^{yx}_\pm - \p_yG^{xx}_\pm\right]\frac{d_l}{d},
\end{equation}
where $S^l_\pm=\langle\hat\sigma^l\rangle_\pm = \pm\frac{d_l}{d}$.



Also, from Eq. (\ref{BC-term}) the BC-dependent $\kv$-resolved spin conductivity for the two-level system  is given by (see appendix(\ref{app-two-level})) 
\begin{equation}\label{BC-term-2level}
    \Pi^{l,(\rm BC)}_{y,\pm} =\mp\frac{e^2}{4 d^3}\Omega^{yx}_\pm [(\p_x\dv)\times\dv]_l.
\end{equation}
Since the electric field direction is fixed at $x$, the indices $b$ and $c$ have been suppressed for simplicity.

\section{Surface states of 3D topological insulators}\label{sec:surface-states}
Materials like $\rm{Bi_2Te_3}$ and $\rm{Bi_2Se_3}$ are time-reversal invariant three-dimensional topological insulators. They are characterized by their helical edge states, protected by a $\mathbb{Z}_2$ invariant\cite{Shen2017}. 
Recent studies \cite{Awadesh,TanayNag} on the surface states of these topological insulators have revealed that the leading order of the anomalous Hall current on the surface is third-order. This provides an opportunity to generate `pure' spin current up to second-order in the applied electric field. 
On the [111] surface of $\rm{Bi_2Te_3}$, the symmetry of the crystal is given by $C_{3v}$, which consists of a three-fold rotation symmetry $C_3$ and a mirror symmetry\cite{Fu-hexagonal-warping}.
The effective low energy Hamiltonian, used to describe the surface states of $\rm Bi_2Te_3$ is given by\cite{Zhang-Nature,model-Hamiltonian-SCZhang}
\begin{equation}\label{ham}
    H(\kv) = v_0(k_x\sigma_y-k_y\sigma_x) +\frac{\lambda}{2}(k_+^3+k_-^3)\sigma_z.
\end{equation}
Here $\lambda$ denotes the strength of the hexagonal warping which appears the first symmetry-allowed correction to the Dirac Hamiltonian\cite{Fu-hexagonal-warping}, and $k_{\pm} = k_x\pm ik_y$.
Eq. (\ref{ham}) can also be written as $H(\kv)= \dv(\kv)\cdot{\bm\sigma}$, where
\begin{equation}\label{d-comp}
    d_x = -v_0k_y,\:\:\:\:\: d_y=v_0k_x,\:\:\:\:\: d_z=\lambda(k_x^3-3k_xk_y^2).
\end{equation}
Transforming to polar coordinates $(k,\phi)$, the energy dispersion is given by
\begin{equation}\label{dispersion}
    \varepsilon_\pm(\kv) = \pm\sqrt{v_0^2k^2+\lambda^2k^6\cos^2(3\phi)}.
\end{equation}

\begin{figure}
\centering
\includegraphics[trim=0cm 0cm 0cm 0cm, clip, width=\linewidth]{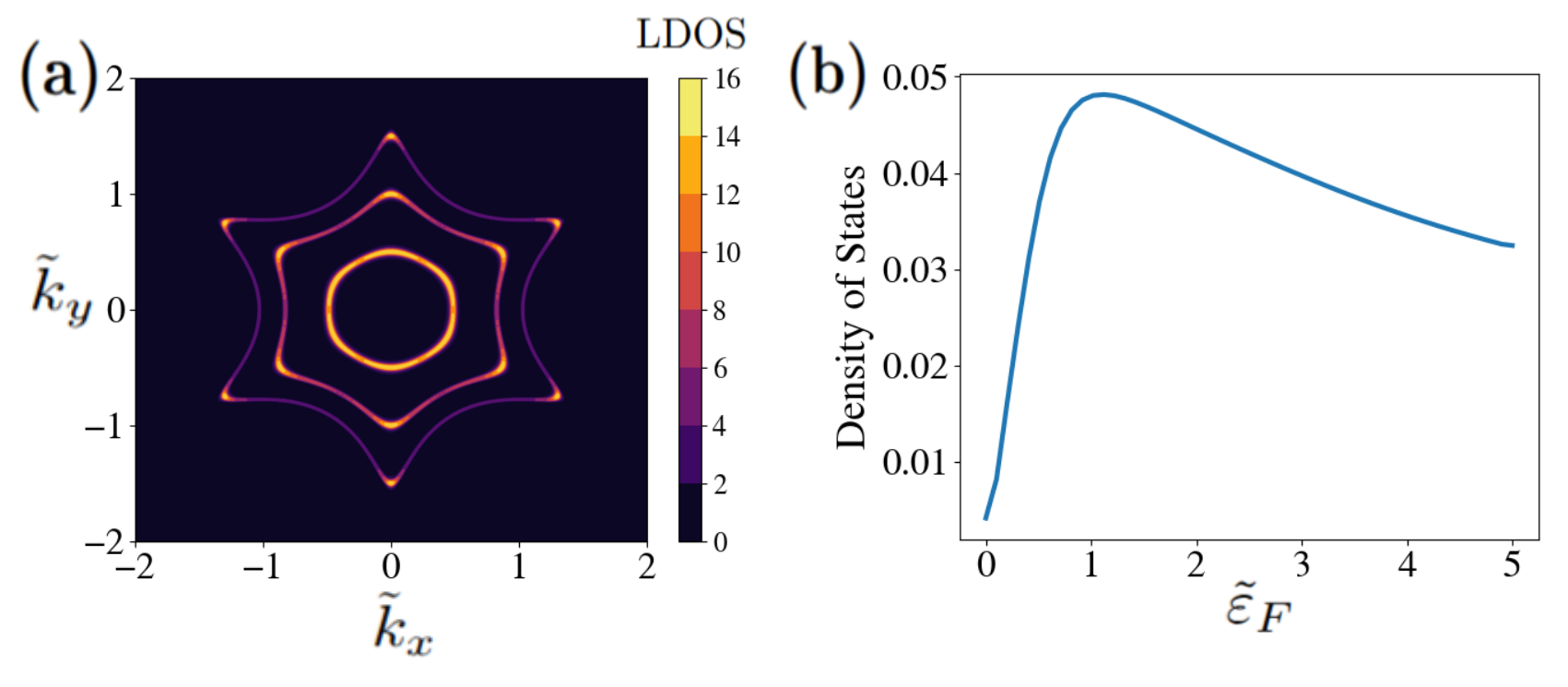}
\captionsetup{format=plain, font=small, labelfont=bf, justification = raggedright}
\caption{(a) The Fermi contours corresponding to three Fermi energies, $\tilde\varepsilon_F = 0.5, 1.0, 1.5$, showing a gradual distortion from circular to snowflake-like with increasing Fermi energy. The LDOS at $\tilde\varepsilon_F = 0.5$ is uniformly distributed along the contour, while on increasing Fermi energy it gets concentrated at the tips of the snowflake contour. (b) The total density of states as a function of Fermi energy.}
\label{fig:Fermi-contour}
\end{figure}

To facilitate the numerical calculation of the spin conductivities, a dimensionless wave-vector is defined as $\tilde k=k/k_0$ where $k_0 = \sqrt{v_0/\lambda}$ is chosen as the typical inverse length scale\cite{Fu-hexagonal-warping}. 
Consequently, the energy dispersion gets scaled by $\varepsilon_0 = v_0\sqrt{v_0/\lambda}$. The typical values of $\lambda$ and $v_0$ are 250 $\mathrm {eV\mathring{A}^3}$ and 2.55 $\mathrm{eV\mathring{A}}$, respectively\cite{Fu-hexagonal-warping}. This sets $k_0 = 0.1\:\mathrm{\mathring{A}^{-1}}$ and $\varepsilon_0 = 0.255\:\mathrm{eV}$.

The Fermi wave-vectors ($k_F$) corresponding to different Fermi energies are computed numerically using Eq. (\ref{dispersion}) and the resulting Fermi contours are plotted in Fig. (\ref{fig:Fermi-contour}(a)).
The effect of warping becomes evident on increasing the Fermi energy, as the isotropy of the energy contours reduces to a six-fold rotation symmetry. The mirror symmetries of the dispersion about the $k_x$ and $k_y$ axes are also visible in Fig. (\ref{fig:Fermi-contour}(a)).
The density of states plotted in Fig. (\ref{fig:Fermi-contour}(b)) shows a sharp initial rise with the Fermi energy and then starts to decrease beyond $\tilde\varepsilon_F = 1$. This is because of the competing k-linear and k-cubic terms in the dispersion equation(\ref{dispersion}) which tend to increase and decrease the density of states, respectively. 
Further, the local density of states (LDOS) in the $\kv$-space becomes anisotropic as the Fermi contours distort into a snow-flake-like shape at higher Fermi energies. Then, the LDOS becomes concentrated only at the tips of the snow-flake contour.

\begin{figure}
\centering
\includegraphics[width=\linewidth]{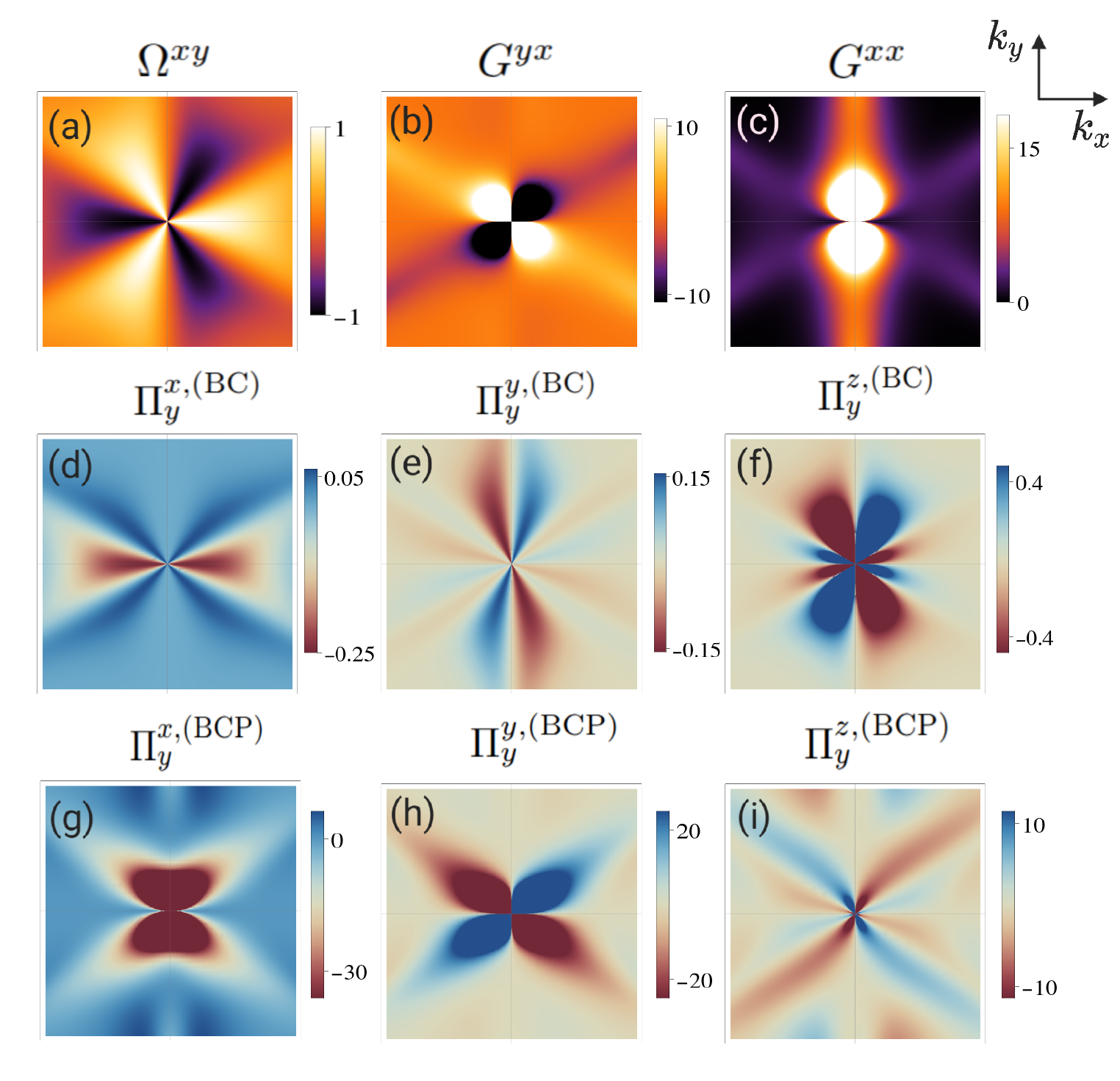}
\captionsetup{format=plain, font=small, labelfont=bf, justification = raggedright}
\caption{ {\bf Top panel}: The momentum space distribution of {\bf (a)} Berry curvature (in units of $\lambda/v_0$), {\bf (b)} transverse and {\bf (c)} longitudinal components of BCP (in units of ${v_0^2}/{16\varepsilon_0^3}$).
{\bf Middle panel}: The three spin components of the BC-induced SOSC (in units of ${e^2\lambda^2}/{4 v_0^3}$). {\bf Bottom panel}: The three spin components of the BCP dipole-induced SOSC (in units of ${e^2\lambda^2}/{16 v_0^3}$) .
Here $k_x$ and $k_y$ are in units of $k_0$ and range from $-1$ to $1$.}
\label{fig:geometric-quantities}
\end{figure}

Using equations (\ref{d-comp}) and (\ref{BC}) the Berry curvature for this system is obtained as,
\begin{equation}
    \Omega^{xy}_{\pm} = \pm \frac{\lambda}{v_0}\frac{\cos3\phi}{(1 + \tilde k^4\cos^23\phi)^{3/2}}.
\end{equation}
It remains finite at each $\kv$-point, including $\kv=0$ where $\Omega^{xy}_{\pm}(\kv\rightarrow 0)=\pm \lambda/v_0$.
The Berry curvature is directly proportional to the hexagonal warping strength which, in turn, plays a crucial role in the first and second-order spin currents.  
It is anisotropic in the $\kv$-space, inheriting its $C_3$ symmetry from the hexagonal warping term but it breaks the mirror symmetry about $k_y$ (\ref{fig:geometric-quantities}(a)). 

The transverse and longitudinal components of the BCP tensor are obtained using equations (\ref{BCP}) and (\ref{d-comp}) as,
\begin{align}
    G^{yx}_\pm =& -\frac{2v_0^2}{16\varepsilon_0^3\tilde d^3}\bigg[\frac{9}{2}\tilde k^4\sin(4\phi) +\frac{1}{2\tilde d^2}\Big\{\tilde k^2\sin(2\phi) \notag\\
    &-6\tilde k^6\sin\phi\cos(3\phi)-9\tilde k^{10}\sin(4\phi)\cos^2(3\phi)\Big\}\bigg], \\
    G^{xx}_\pm =& \frac{v_0^2}{16\varepsilon_0^3\tilde d^3}\bigg[1+9\tilde k^4\cos^2(2\phi) \notag\\
    &- \frac{1}{\tilde d^2}\left\{\tilde k\cos\phi + 3\tilde k^5\cos(2\phi)\cos(3\phi)\right\}^2\bigg],
\end{align}
where $\tilde d = \sqrt{\tilde k^2 + \tilde k^6\cos^2(3\phi)}$.
They have been plotted in Fig. (\ref{fig:geometric-quantities}(b),(c)). $G_{yx}$ shows a quadrupolar distribution, odd under the mirror symmetries about both axes while $G_{xx}$ is dipolar and even under the mirror symmetries\cite{TanayNag,Awadesh}. Neither of the two BCP components show a six-fold rotation symmetry like the Berry curvature. 

\subsection{Linear spin current}

Although the Berry curvature is finite everywhere in the $\kv$-space, time-reversal symmetry enforces $\Omega^{xy}(-\kv) = -\Omega^{xy}(\kv)$ and makes its integral over the $\kv$-space vanish.
Thus, there is no first-order anomalous Hall effect in $\mathcal{T}$-preserved systems. This, however, changes for spin Hall effect.
The first-order spin conductivity is given by Eq. (\ref{1st-order-cond}), which can be rewritten as,
\begin{equation}\label{linear-cond}
    \chi^l_{ab} = \frac{e}{2}\sum_n\int_\kv {\Xi}^{l}_{ab,n} f^{(0)}_n(\kv),
\end{equation}
where $\Xi^l_{ab}(\kv) =\Omega^{ab}_nS^l_n$ is the spin-weighted Berry curvature.
Using Eq. (\ref{d-comp}), $\Xi^l_{ab}(\kv)$ is obtained for $l=x,y,z$ as 
\begin{align}\label{sbc1}
    &\Xi^x_{yx}(\kv) = -\frac{\lambda}{v_0} \frac{\tilde k^4\cos(3\phi)\sin\phi}{\tilde d^4}, \\
    &\Xi^y_{yx}(\kv) = -\frac{\lambda}{v_0} \frac{\tilde k^4\cos(3\phi)\cos\phi}{\tilde d^4},\label{sbc2} \\
    &\Xi^z_{yx}(\kv) = -\frac{\lambda}{v_0} \frac{\tilde k^6\cos^2(3\phi)}{\tilde d^4}.\label{sbc3}  
\end{align}

\begin{figure}
\centering
\includegraphics[width=\linewidth]{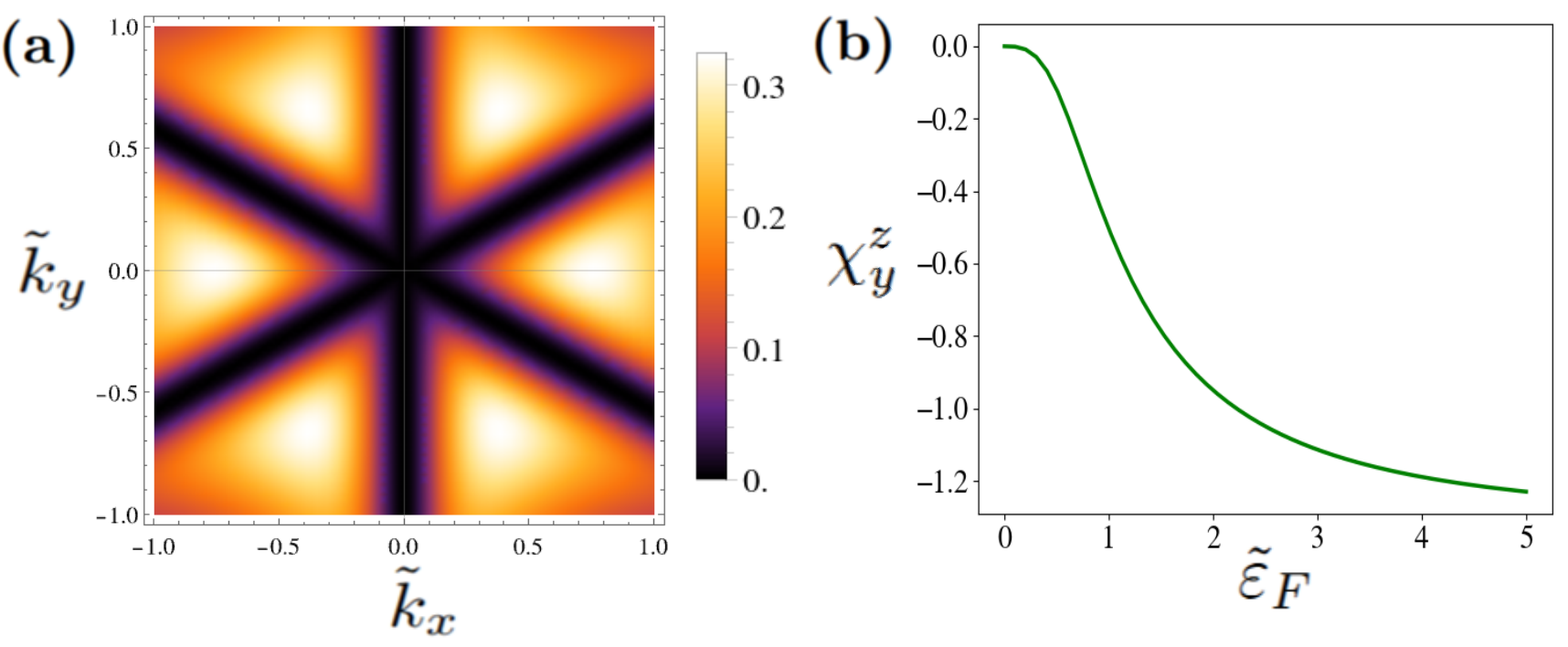}
\captionsetup{format=plain, font=small, labelfont=bf, justification = raggedright}
\caption{{\bf (a)} Momentum space distribution of the spin $z$-component of spin-weighted Berry curvature, $\Xi^z_{yx}$ (in units of $-\lambda/v_0$. {\bf (b)} Linear spin Hall conductivity (in units of $e/2$), with out-of-plane spin polarization as a function of Fermi energy.}
\label{fig:1st-order-cond}
\end{figure}

Out of these three components, the out-of-plane spin component $\Xi^z_{yx,\pm}$ turns out to be the only non-vanishing component from Eq. (\ref{linear-cond}). The in-plane components vanish when integrated over the k-space.
However, unlike the spin Berry curvature in 2D Rashba systems, the linear in-plane components here, are not prohibited due to $\mathcal{T}$ alone. In fact, equations (\ref{sbc1}-\ref{sbc3}) shows that the spin-weighted Berry curvature is finite at all $\kv$-points, for all the three spin components.
The cancellation of the linear in-plane components, in this case, is due to the anisotropy band dispersion, resulting from the $C_3$ symmetry of the material which propagates to the Berry curvature and spin polarization.    
The linear spin Hall conductivity is plotted as a function of Fermi energy in Fig. (\ref{fig:1st-order-cond}(b)).

\subsection{Second-order spin current}

The second-order spin conductivity (SOSC), as given in Eq. (\ref{2nd-order-spin-cond}) can be classified as extrinsic and intrinsic based on their dependence on the scattering time-scale $\tau$.

\begin{figure}
\centering
\includegraphics[width=\linewidth]{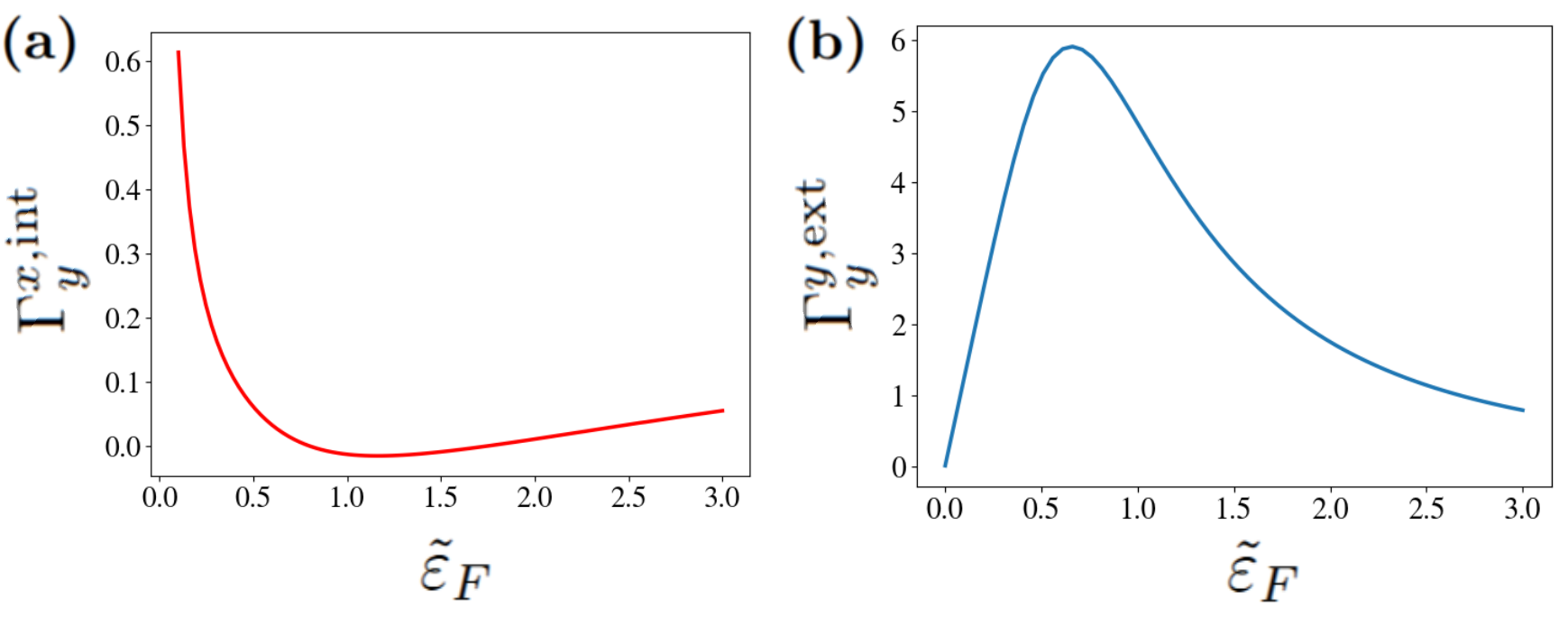}
\captionsetup{format=plain, font=small, labelfont=bf, justification = raggedright}
\caption{{\bf (a)} Intrinsic SOSC, with spins polarized in the direction of the applied electric field (in units of $10^2\Gamma_0$), where $\Gamma_0=e^2\lambda/v_0^2$). {\bf (b)} Extrinsic SOSC with spin polarization along the direction of the spin current, (in units of $10^{-3}\Gamma_0$) as a function of Fermi energy. Here $\tau = 1 $ ps.}
\label{fig:2nd-order-cond}
\end{figure}

The three spin components of the $\kv$-resolved BCP dipole-induced SOSC ($\Pi^{l,\rm{BCP}}_{y,\pm}$) have been calculated using Eq. (\ref{bcp-term-2-level}) and plotted in Fig. (\ref{fig:geometric-quantities}(g) - \ref{fig:geometric-quantities}(i)).
Similarly, the $\kv$-resolved BC-induced SOSC ($\Pi^{l,\rm{BC}}_{y,\pm}$) is calculated using Eq. (\ref{BC-term-2level}) and plotted in Fig. (\ref{fig:geometric-quantities}(d) - \ref{fig:geometric-quantities}(f)).
Together, they constitute the intrinsic second-order spin Hall response.
In both cases, out of the three spin components, only the $x$-component survives the integration over the $\kv$-space.
It can be clearly seen in the figures (\ref{fig:geometric-quantities}(d)-(i)) that $\Pi^{x,\rm{BC}}_{y,\pm}$ and $\Pi^{x,\rm{BCP}}_{y,\pm}$ have mirror symmetries with respect to the $k_x$ and $k_y$ axes. The other components are anti-symmetric about the $k_x$ and $k_y$ axes which makes their integrals vanish.  
This is once again attributed to the anisotropy induced by the $C_3$ symmetry in the $\kv$-space distribution of the various components of the SOSC.
The $x$-polarized spin current can be classified as a field-polarized spin current, since the spins are polarized in the direction of the applied electric field and perpendicular to the direction of transport.
The total intrinsic spin conductivity, $\Gamma^{x,\rm int}_{y} = \Gamma^{x,\rm{BC}}_{y} + \Gamma^{x,\rm{BCP}}_{y}$ is plotted as a function of Fermi energy in Fig. (\ref{fig:2nd-order-cond}(a)).

The extrinsic second-order spin current is due to the dipole of the spin-weighted Berry curvature $\Xi^l_{ab}(\kv)$,
\begin{equation}\label{extrinsic}
    \Gamma^{l, \rm ext}_{abc} = \frac{e^2\tau}{2\hbar}\sum_n \int_\kv \Xi^{l}_{ab,n}\frac{\p f^{(0)}_n}{\p k_c}. 
\end{equation}
The derivative when acting on the Fermi distribution yields a Dirac-delta function at $k_F$, making the extrinsic SOSC a Fermi surface effect.
Now, anisotropy in the Berry curvature distribution is essential to obtain its dipole moment. In this case, however, despite having anisotropy, the Berry curvature dipole (BCD) vanishes due to the six-fold rotation symmetry of the hexagonal warping term.
This leaves an opportunity for the spin BCD to generate pure extrinsic spin current in the second-order.
The calculation of the extrinsic SOSC is shown in appendix(\ref{app:ext}) and its dominant in-plane spin component is plotted in Fig. (\ref{fig:2nd-order-cond}(b)).
The $x$-component almost vanishes, whereas the $y$ component shows a peak at smaller Fermi energies and decay smoothly on increasing Fermi energy. This is because the integral is evaluated at the Fermi surface and does not accumulate the contribution of the states at lower energies. Thus, when $\tilde\varepsilon_F(k_F)$ is increased, the higher powers of $k_F$ appearing in the denominator of the SOSC expression (see appendix(\ref{app:ext})) due to hexagonal warping suppresses the integral numerically.
Moreover, the decrease in density of states at higher Fermi energy also contributes to the suppression of the SOSC. 
The $y$-polarized in-plane spin current is classified as a collinearly polarized spin current (CPSC) having the spin polarized in the same direction as the transport and may have important application in field-free magnetization switching\cite{CPSC-wang, SOT-MRAM}.

In Fig. (\ref{fig:2nd-order-cond}) the unit of SOSC is given in terms of $ \Gamma_0 = e^2\lambda/v_0^2 $. Converting to real units, $\Gamma^{x, \rm int}_y$ turns out to be close to $0.4 \:e\:\mu m/V$, which is of the same order as the intrinsic second-order spin conductivity, reported in the 2D Rashba-Dresselhaus electron gas.
Similar to the second-order spin conductivity in 2D hole gas\cite{our-paper}, we find that the intrinsic response is much larger in magnitude than the extrinsic one.
These second-order spin currents can be probed in various ways, including the magneto-optic Kerr effect (MOKE) that detects spin accumulation at the sample edges\cite{SHE_Kato_exp, SHE-expt-semiconductor, SHE-expt-heavy-metal},
spin-to-charge conversion using the inverse spin Hall effect and detecting the second harmonic response of the applied ac electric field\cite{Saitoh, 2nd-harmonic-spin-current}, 
and the detection of spin-orbit torque exerted by the spin currents when injected into a magnetic material\cite{ST-FMR, spin-torque-switching, Garello_magnetization-switching}.

\section{Discussion and Conclusion}\label{sec:discussion}
In summary, this article explored SHE in the surface states of topological insulator ($\rm{Bi_2Te_3}$), in the presence of hexagonal warping. 
Compared to the non-linear anomalous Hall effect, the effect of hexagonal warping on spin transport is more profound. 
The Berry curvature that leads to the linear spin Hall current is contributed solely by the hexagonal warping. It also leads to anisotropy in the band dispersion and the $\kv$-space distributions of various geometric quantities.
The overall anisotropy is an artefact of the time-reversal and $C_3$ symmetries of the system which determine the fate of the various spin components of the linear and non-linear conductivities.

Our results reveal that the linear spin current is purely out-of-plane, while the in-plane spin components appear only in the second-order response. 
The intrinsic SOSC is contributed by the spin-dependent Berry curvature and BCP dipole both of which lead to a field-polarized spin current.
The extrinsic SOSC is a Fermi surface effect coming from the spin-dependent BCD and leads to a collinearly polarized spin current.
Thus the band anisotropy, which in this case is controlled by the crystal symmetry, can be harnessed as a selection switch for the different components of the spin conductivity. 
It will be interesting to study the effect of strain or other perturbations that modify the crystal symmetry. It may lead to a different anisotropy which may unlock the other components of the spin conductivity as well.
\begin{center}
    \textbf{DATA AVAILABILITY STATEMENT}\\
\end{center}
The data that support the findings of this study are available from the corresponding author upon reasonable request.\\
\begin{center}
    \textbf{ ACKNOWLEDGEMENT}\\
\end{center}
SC acknowledges IIT Kanpur for providing Ph.D. fellowship.

\appendix

\section{Berry curvature-induced second-order spin current}\label{app-two-level}
From Eq. (\ref{BC-term}) the $\kv$-resolved BC-dependent SOSC for a two-level system is given by
\begin{equation}
 \Pi^{l,(\rm BC)}_{abc,\pm} = e^2\Omega^{ab}_\pm\:\Im\left\{\frac{\langle \mp\vert\hat \sigma^l\vert\pm\rangle\langle\pm\vert \hat{v}^c\vert\mp\rangle}{4d^2}\right\},   
\end{equation}
where the band velocity operator can be written using Eq. (\ref{2level-Ham}), as
\begin{equation}
    \hat{ v}^c = \p_c(d_i\sigma^i)=(\p_c d_i)\sigma^i.
\end{equation}
Then,
\begin{equation*}
    \langle\pm\vert \hat{v}^c\vert\mp\rangle = (\p_cd_i)\langle\pm\vert\sigma^i\vert\mp\rangle,
\end{equation*} and
\begin{equation*}
    \Im \left\{\langle \mp\vert\hat \sigma^l\vert\pm\rangle\langle\pm\vert \hat{v}^c\vert\mp\rangle\right\} = (\p_cd_i)\:\Im\left\{\langle\mp\vert\hat \sigma^l\vert\pm\rangle \langle\pm\vert\hat\sigma^i\vert\mp\rangle \right\}
\end{equation*}
On simplification, this reduces to
\begin{align*}
   \Im &\left\{\langle \mp\vert\hat \sigma^l\vert\pm\rangle\langle\pm\vert \hat{v}^c\vert\mp\rangle\right\}  \\
   &=(\p_cd_i)\:\Im\{\delta_{il} - d_id_l/d^2 \pm i\epsilon_{ilj}d_j/d\}\\
   &=\pm \epsilon_{ilj}(\p_cd_i)d_j/d\\
   &=\mp[(\p_c\dv)\times\dv]_l.
\end{align*}
Hence, Eq. (\ref{BC-term-2level}) becomes,
\begin{equation}
    \Pi^{l,(\rm BC)}_{abc,\pm} =\mp\frac{e^2}{4 d^3}\Omega^{ab}_\pm [(\p_c\dv)\times\dv]_l.
\end{equation}
Again, fixing the electric field along the $x$-direction, we get
\begin{equation}
    \Pi^{l,(\rm BC)}_{y,\pm} =\mp\frac{e^2}{4 d^3}\Omega^{yx}_\pm [(\p_x\dv)\times\dv]_l.
\end{equation}
Their explicit forms are as follows:
\begin{align}
    &\Pi^{x,\rm BC}_{y,\pm} = -\frac{e^2\lambda^2}{4 v_0^4 \tilde d^6}\tilde k^6 \cos(3\phi)\cos^3\phi, \\
    &\Pi^{y,\rm BC}_{y,\pm} = \frac{3e^2\lambda^2}{4 v_0^4 \tilde d^6}\tilde k^6 \cos(3\phi)\cos(2\phi)\sin\phi,\\
    &\Pi^{z,\rm BC}_{y,\pm} = -\frac{e^2\lambda^2}{4 v_0^4 \tilde d^6}\tilde k^4 \cos(3\phi)\sin\phi.
\end{align}
Their $\kv$-space distributions have been shown in Fig. (\ref{fig:geometric-quantities}(d)-\ref{fig:geometric-quantities}(f)).

\section{Extrinsic second-order spin current}\label{app:ext}
Within the relaxation time approximation of the Boltzmann transport equation, the extrinsic contribution to the SOSC is given by Eq. (\ref{extrinsic}) as
\begin{equation}
    \Gamma^{l,ext}_{y} = \frac{e^2\tau}{2\hbar}\sum_n \int_\kv \Xi^{l}_{yx,n}\frac{\p f^{(0)}_n}{\p k_x}. 
\end{equation}
In the low temperature limit ($T\rightarrow0$) the derivative of the Fermi distribution is given by,
\begin{equation*}
    \frac{\p f^{(0)}}{\p\varepsilon_\pm} = -\delta(\varepsilon_\pm -\mu) = -\frac{\delta(k-k_F^\pm)}{\left\vert\frac{\p \varepsilon_F^\pm}{\p k_F^\pm}\right\vert}.
\end{equation*}
Now, the extrinsic SOSC becomes
\begin{equation}\label{extrinsic2}
    \Gamma^{l,ext}_{y,\pm} = -\frac{e^2\tau}{2\hbar}\int_\kv\Xi^l_{yx,\pm}\frac{\p \varepsilon_\pm}{\p k_x}\frac{\delta(k-k_F^\pm)}{\left\vert\frac{\p \varepsilon_F^\pm}{\p k_F^\pm}\right\vert},
\end{equation}
where $\frac{\p \varepsilon_\pm}{\p k_x} = \pm \left(kv_0^2\cos\phi + 3\lambda^2k^3\cos(3\phi)\cos(2\phi)\right)/d$ and $\frac{\p \varepsilon_F^\pm}{\p k_x} = \pm \left(v_0^2k^\pm_F + 3(k_F^\pm)^5\cos^2(3\phi)\right)/d$.
On simplification Eq. (\ref{extrinsic2}) becomes
\begin{align*}
     \Gamma^{l,ext}_{y,\pm} = &\mp\frac{e^2\tau}{2\hbar}\int_0^{2\pi}d\phi\: k_F^\pm \Xi^l_{yx,\pm}(k_F^\pm)\\
     &\frac{\left( v_0^2\cos\phi + 3\lambda^2(k^\pm_F)^4\cos(3\phi)\cos(2\phi)\right)}{\vert v_0^2 + 3(k_F^\pm)^4\cos^2(3\phi) \vert}.
\end{align*}
For a given Fermi energy, $k_F$ is a function of $\phi$. With increase in the Fermi energy, $k_F$ also increases numerically which explains the rise in the spin conductivity (Fig. (\ref{fig:2nd-order-cond}(b))). However, when $k_F>1$ (in units of $k_0$) the higher powers of $k_F$ dominate in the denominator of $\Xi^l_{yx}$ in the above integrand which reduces the overall magnitude of the spin conductivity.

\bibliography{bibliography}

\end{document}